\documentclass[aps, prd, reprint, 11pt, notitlepage, a4paper, floats, floaqutfix, amsmath, amssymb, amsfonts, superscriptaddress, showpacs, showkeys, nofootinbib, longbibliography]{revtex4-2}

\pdfoutput=1

\usepackage{rotating}
\usepackage{epsfig}
\usepackage{graphics}
\usepackage{graphicx}
\usepackage{amsmath,amsfonts,amssymb,amsthm}
\usepackage[usenames,dvipsnames]{xcolor}
\usepackage{wasysym}
\usepackage{times}
\usepackage{mathptmx}
\usepackage{gensymb}
\usepackage{appendix}
\usepackage{listings}
\usepackage{url}
\usepackage[normalem]{ulem}
\usepackage{alltt}
\usepackage[colorlinks, allcolors=dodgerblue]{hyperref}
\usepackage{cleveref}
\usepackage{longtable}
\usepackage{enumitem}
\setlist{nosep}
\usepackage{color}
\usepackage{calc}
\usepackage{tensor}
\usepackage{bm}
\usepackage{times}
\usepackage{multirow}
\usepackage[varg]{txfonts}
\usepackage{float}
\usepackage{dcolumn}
\usepackage[nolist,nohyperlinks]{acronym}
\usepackage{xspace}
\usepackage[english]{babel}
\usepackage[abs]{overpic}
\usepackage{pict2e}
\usepackage[caption=false]{subfig}
\allowdisplaybreaks[1]
\usepackage[utf8]{inputenc}
\usepackage{gensymb}
\usepackage{bm}
\usepackage{stackengine}
\usepackage{boldline,multirow}
\usepackage{braket}
\usepackage{longtable}
\usepackage{booktabs}
\usepackage{tabularx}
\usepackage{rotating}
\usepackage{listings}
\usepackage{courier}
\usepackage{setspace}
\usepackage{accsupp}
\usepackage{rotating}
\usepackage{siunitx}

\newcommand{\emptyaccsupp}[1]{\BeginAccSupp{ActualText={}}#1\EndAccSupp{}}

\graphicspath{{pics/}}

\newcommand{\AEI}{Max Planck Institute for Gravitational Physics~(Albert~Einstein~Institute), Am M\"uhlenberg 1, Potsdam 14476, Germany}

\newcommand{\Maryland}{Department of Physics, University of Maryland, College Park, MD 20742, USA}
\newcommand{\URI}{Department of Physics, East Hall, University of Rhode Island, Kingston, RI 02881, USA}
\newcommand{\URICCR}{Center for Computational Research, Tyler Hall, University of Rhode Island, Kingston, RI 02881, USA}

\definecolor{dodgerblue}{HTML}{1E90FF}
\definecolor{viennared}{HTML}{DA0A14}

\def\dd{\mathrm{d}}

\newcommand{\pyeobnr}{\texttt{pySEOBNR}}

\newcommand{\qq}{\symbol{34}} 

\newcommand{\gw}{\textsc{gw}}

\begin{document}

\title{\pyeobnr: a software package for the next generation\\of effective-one-body multipolar waveform models}

\author{Deyan P. Mihaylov}\email{deyan@aei.mpg.de}
\affiliation{\AEI}
\author{Serguei Ossokine}\email{serguei.ossokine@aei.mpg.de}
\affiliation{\AEI}
\author{Alessandra Buonanno}
\affiliation{\AEI}
\affiliation{\Maryland}
\author{Hector~Estelles}
\affiliation{\AEI}
\author{Lorenzo Pompili}
\affiliation{\AEI}
\author{Michael Pürrer}
\affiliation{\URI}
\affiliation{\URICCR}
\affiliation{\AEI}
\author{Antoni Ramos-Buades}
\affiliation{\AEI}

\date{\today}

\begin{abstract}
    \noindent
    We present \pyeobnr, a Python package for gravitational-wave
    (GW) modeling developed within the effective-one-body (EOB)
    formalism. The package contains an extensive framework to
      generate state-of-the-art inspiral-merger-ringdown waveform
        models for compact-object binaries composed of black holes and
        neutron stars. We document and demonstrate how to use the 
        built-in quasi-circular precessing-spin model \texttt{SEOBNRv5PHM}, 
        whose aligned-spin limit (\texttt{SEOBNRv5HM}) has been calibrated to numerical-relativity simulations
        and the nonspinning sector to gravitational self-force data using \pyeobnr. Furthermore, \pyeobnr~contains the
        infrastructure necessary to construct, calibrate, test, and
        profile new waveform models in the EOB 
        approach. The efficiency and flexibility of \pyeobnr~ will be crucial 
to overcome the data-analysis challenges posed by upcoming and next-generation GW detectors 
on the ground and in space, which will afford the possibility to observe all compact-object binaries in our Universe.
\end{abstract}

\maketitle

\onecolumngrid
\section*{Metadata}

\begin{center}
\begin{table}[h!]
\begin{tabular}{ l l }
\toprule
    Current code version & 0.1 \\ 
\midrule
    Permanent link to code / repository & \href{https://git.ligo.org/waveforms/software/pyseobnr}{\texttt{git.ligo.org/waveforms/software/pyseobnr}} \\
\midrule
    Legal Code License & GNU General Public License \\
\midrule
    Code versioning system used & \texttt{git} \\
\midrule
    Software languages, tools, & \multirow{ 2}{*}{\texttt{Python}, \texttt{Mathematica}} \\
    and services used & \\
\midrule
    Compilation requirements & \multirow{ 2}{*}{\texttt{wheel}, \texttt{setuptools}, \texttt{numpy}} \\
    and dependencies & \\
\midrule
    Link to developer & \multirow{ 2}{*}{\href{https://waveforms.docs.ligo.org/software/pyseobnr}{\texttt{waveforms.docs.ligo.org/software/pyseobnr}}} \\
    documentation / manual & \\
\midrule
    Support email for questions & \href{mailto:pyseobnr@aei.mpg.de}{\texttt{pyseobnr@aei.mpg.de}} \\
\bottomrule
\end{tabular}
\caption{Code metadata}
\label{table:metadata}
\end{table}
\end{center}

\medskip

\section{Motivation for developing \pyeobnr}

\noindent
The field of gravitational-wave (\gw) astronomy has made significant
progress since the first detection of GWs from the merger of a
binary black hole in 2015 \cite{LIGOScientific:2016aoc}. The
first and second observing runs of the LIGO and Virgo ground-based
detectors \cite{TheLIGOScientific:2014jea, TheVirgo:2014hva} resulted
in only a handful of detections \cite{LIGOScientific:2018mvr,
  Venumadhav:2019lyq}, whereas during the third-observing run, over a
hundred events were detected \cite{LIGOScientific:2020ibl,
  LIGOScientific:2021usb, LIGOScientific:2021djp, Nitz:2021uxj,
  Olsen:2022pin} including GWs from a wider variety of compact-object
binaries.  The upcoming fourth-observing run, which will also include the KAGRA detector~\cite{KAGRA:2018plz,KAGRA:2020tym}, planned upgrades to the existing detectors~\cite{KAGRA:2013rdx,LIGOScientific:2018jsj,LIGOScientific:2020kqk} and
next-generation GW detectors on the ground (for example the Einstein
Telescope \cite{Punturo:2010zz} and Cosmic Explorer
~\cite{Reitze:2019iox,Evans:2021gyd}), as well as the space-based detector LISA~\cite{amaro2017laser}, will result in an even
greater detection rate. In order to take full advantage of these
important developments, we need accurate models of the GW signals
emitted by compact-binary coalescences~\cite{Purrer:2019jcp} 
to infer  unique information about gravity, astrophysics, cosmology and fundamental physics.

The golden standard of waveform modeling is performing full numerical-relativity (NR)
simulations~\cite{Pretorius:2005gq, Campanelli:2005dd,
  Baker:2005vv}. Over the past decade, a number of NR waveform
catalogues have been built, using several different software
packages~\cite{Jani:2016wkt, Healy:2017psd, Healy:2019jyf,
  Boyle:2019kee, Healy:2022wdn,Hamilton:2023qkv}. However, their computational cost
makes them prohibitively expensive to carry out across the entire
binary's parameter space and also limits their duration. For this reason,
semi-analytical waveform models which combine analytical predictions for the 
two-body dynamics and GW radiation with
NR data were developed.  Presently, there are a number of waveform
models available that are built using one of several approaches
outlined below.

NR surrogate models are constructed by interpolating a sufficient
number of NR waveforms across the desired parameter
space~\cite{Blackman:2015pia, Blackman:2017dfb, Blackman:2017pcm,
  Varma:2018mmi, Varma:2019csw, Williams:2019vub, Rifat:2019ltp,
  Islam:2021mha, Islam:2022laz, Yoo:2022erv}. This approach has been
successful in accurately modeling systems where higher multipoles are
important~\cite{Varma:2018mmi}, for precessing-spin
binaries~\cite{Blackman:2017pcm, Varma:2019csw} and for eccentric
systems~\cite{Islam:2021mha}. However, NR surrogate waveforms are limited in their
length (unless hybridized to analytical waveforms), and are restricted to those 
regions of parameter space where enough NR waveforms are available.

The other two dominant approaches to waveform modeling are the
phenomenological and effective-one-body (EOB) frameworks.  The
phenomenological waveform models
(\texttt{IMRPhenom})~\cite{Pan:2007nw, Ajith:2007qp, Ajith:2009bn,
  Santamaria:2010yb, Hannam:2013oca, Husa:2015iqa, Khan:2015jqa,
  London:2017bcn, Khan:2018fmp, Khan:2019kot, Dietrich:2019kaq,
  Pratten:2020fqn, Pratten:2020ceb, Garcia-Quiros:2020qpx,
  Estelles:2020osj, Estelles:2020twz, Estelles:2021gvs,
  Hamilton:2021pkf} combine post-Newtonian (PN) and EOB analytic expressions
for the inspiral with NR information for the lead-up to merger and
merger-ringdown portions of the waveform. Their computational
efficiency has made them a standard choice for data analysis in GW
astronomy.  The other waveform family routinely used in GW observations is the 
EOB one, which combines an analytical description of the
two-body dynamics based on PN and small-mass ratio perturbation theory with NR
data for the strong-field regime~\cite{Buonanno:1998gg,
  Buonanno:2000ef, Damour:2000we, Damour:2001tu, Buonanno:2005xu}. EOB
models have been developed for a broad range of systems: non-spinning,
aligned-spin, precessing-spin, and eccentric binaries.  The two most
notable EOB families are \texttt{SEOBNR}~\cite{Bohe:2016gbl,
  Cotesta:2018fcv, Ossokine:2020kjp, Cotesta:2020qhw,
  Ramos-Buades:2021adz, Mihaylov:2021bpf} and
\texttt{TEOBResumS}~\cite{Nagar:2018zoe, Nagar:2019wds, Nagar:2020pcj,
  Gamba:2021ydi, Riemenschneider:2021ppj, Chiaramello:2020ehz}. In
this work we will focus on the \texttt{SEOBNR} family, which
has been employed by the LIGO-Virgo-KAGRA (LVK) Collaboration to analyse GW signals of
compact-binary coalescences, infer their astrophysical properties, and
test the theory of General Relativity~(GR).

While waveforms from the \texttt{SEOBNR} family have been widely used for GW
analyses due to their high accuracy, they are often computationally expensive with nested sampling and MCMC methods~\cite{Skilling:2006gxv,Veitch:2014wba,Ashton:2018jfp}. 
This issue has been tackled in several ways. 
One possible approach is to construct surrogate models~\cite{PhysRevD.103.064015, PhysRevD.106.104029, Field:2013cfa, Purrer:2014fza, Purrer:2015tud, Blackman:2015pia, Blackman:2017pcm, Blackman:2017dfb, Lackey:2018zvw, Doctor:2017csx, Setyawati:2019xzw, Varma:2018mmi, Varma:2019csw, Cotesta:2020qhw, Gadre:2022sed, Thomas:2022rmc}. 
Another possibility is to use alternative samplers for parameter estimation, such as \texttt{RIFT}~\cite{Lange:2018pyp}, \texttt{parallel Bilby}~\cite{Smith:2019ucc}, \texttt{VItamin\_C}~\cite{Gabbard:2019rde}, \texttt{VARAHA}~\cite{Tiwari:2023mzf} or others, which have been developed with the goal of decreasing the walltime of analyses.
Notably, \texttt{DINGO}~\cite{Green:2020dnx, Dax:2021tsq, Dax:2022pxd}, a software package based on machine learning, has also been used for inference studies with \texttt{SEOBNR}. 
However, some events are still extremely challenging to analyse and those approaches alone do not readily scale to a large number of events. 
The construction of reduced-order or surrogate models to accelerate EOB waveform generation~\cite{Field:2013cfa, Purrer:2014fza, Purrer:2015tud, Lackey:2016krb, Lackey:2018zvw, Cotesta:2020qhw, Gadre:2022sed, Tissino:2022thn, Khan:2020fso, Thomas:2022rmc} can be challenging and time-consuming, thus limiting the pace with
which new advances can be incorporated into the \texttt{SEOBNR} models.
All these factors underscore the need to build a new framework for rapid development and generation of new models.

The need to build more accurate waveform models is also driven by the increasing sensitivity of GW detectors.  Upcoming observations of the LVK Collaboration, as well as future detectors like LISA, Einstein Telescope and Cosmic Explorer, will provide the opportunity for pioneering discoveries about the nature and origins of compact objects and gravity, fundamental 
physics and cosmology, provided we are equipped with high-precision and efficient gravitational waveforms, which include all 
physical effects and can be used in such studies. For example,  using accurate and computationally efficient waveform 
models for binary neutron stars (BNSs) and neutron-star--black-hole (NSBH) binaries, we would be able to probe the internal structure and equation of state of neutron stars, as well as characterize their population and origin, including details about the low-end mass gap of compact objects.

The higher sensitivity of future detectors poses another challenge. In
particular, most of these detectors will observe in a lower/wider
frequency band, resulting in much longer signals.  This necessitates
significantly more computationally efficient waveform models. Thus, a
combination of Fourier-domain and time-domain modeling should be
explored, as well as the use of novel hardware architectures,
e.g. GPUs and parallel computing. This in turn requires a new
infrastructure and framework for waveform development.

In order to meet the need for faster, more accurate and more
physically complete GW models, we have developed a completely new code
base, \pyeobnr, which implements all tools necessary for the development of 
the next generation of \texttt{SEOBNR} waveform models in a modern and easily maintainable environment. 
On one hand, the software package provides access to the aligned-spin
\texttt{SEOBNRv5HM}~\cite{Pompiliv5} and precessing-spin
\texttt{SEOBNRv5PHM} models~\cite{RamosBuadesv5} for use in scientific applications.  For example,  several different parameter estimation codes are being adapted to use these models, such as \texttt{bilby}~\cite{Ashton:2018jfp,
  Romero-Shaw:2020owr}, \texttt{DINGO} and \texttt{RIFT}. On the other hand,
\pyeobnr~provides a suite of tools for the user to develop new models.

Technical information about \pyeobnr~can be found in
Table~\ref{table:metadata}. We choose
\texttt{python}~\cite{pub:18204, van1995python, 10.5555/1593511} to build the framework, due to
its rich scientific ecosystem, maintainability and wide adoption in
the GW community ~\cite{alex_nitz_2023_7692098, pyring,
  Carullo:2019flw, Ashton:2018jfp, Romero-Shaw:2020owr}. In addition,
certain parts of the code are built with
\texttt{cython}~\cite{behnel2011cython}, a compiled extension to
\texttt{python}, in order to achieve better efficiency where
necessary, while retaining a lot of flexibility. Theoretical results,
which are the foundation of the \texttt{SEOBNR} family of models, can
be provided via \texttt{Mathematica}~\cite{mathematica} notebooks, which are
then parsed and used by \pyeobnr. The package is published under the
GNU General Public License~\cite{gpl}. Version control is provided
through \texttt{git}~\cite{git,chacon2014pro}, which also offers a
platform for streamlined management of code development, issue
tracking, and publishing updated releases of the package. The most
recent version of \pyeobnr~is available
through the linked \texttt{git} repository. Stable versions of \pyeobnr~ are
published through the Python Package Index (PyPI)~\cite{pypi} repository (a user may install the 
latest stable version via \texttt{pip~install~pyseobnr}). In order to facilitate the
broader adoption of this new software package, the authors welcome
questions, suggestions, and bug reports at the provided electronic
mail address.

In this article we review the current state-of-the-art of the
\pyeobnr~package. In Section~\ref{sec:package} we demonstrate its
functionality, we describe the necessary inputs and possible options, and
we explain how the submodules interact with each other. In
Section~\ref{sec:UseCases} we provide code samples that showcase the
main use cases of \pyeobnr. Finally, in Section~\ref{sec:future}, we
conclude by providing a roadmap for future developments. This article
does not represent a complete record of the various functionalities of
\pyeobnr. For further information about the package, please refer to the full documentation,
linked in Table~\ref{table:metadata}. 
The theoretical results on which the waveform models are based can be found in Ref.~\cite{Khalilv5}. 
For details on the specifics of the models please consult Refs.~\cite{Pompiliv5, VandeMeentv5} (aligned-spin model \texttt{SEOBNRv5HM} and calibration pipelines), Ref.~\cite{RamosBuadesv5} (precessing-spin model \texttt{SEOBNRv5PHM}) and references therein.

\section{Description of the software package}
\label{sec:package}

\noindent
The user interface of \pyeobnr~is implemented in \texttt{python} \cite{10.5555/1593511} (more precisely, the package is compatible with Python v3.8 and above) in order to facilitate ease of use and quick adoption by the GW community. 
The package depends on a number of established, well-known, and regularly maintained packages under open-software licenses: \texttt{numpy} \cite{harris2020array}, \texttt{scipy} \cite{2020SciPy-NMeth}, \texttt{cython} \cite{behnel2011cython}, \texttt{GSL}~\cite{gsl, gough2009gnu} , \texttt{pygsl}~\cite{pygsl} as well as \texttt{lal} and \texttt{lalsimulation}~\cite{lalsuite, swiglal} and others are used in the extensions for waveform generation. 
In addition, \texttt{wolframclient}~\cite{wolframclient} and \texttt{bilby}~\cite{Ashton_2019} are used for the extensions that allow the development and calibration of new waveform models.
Table~\ref{table:metadata} contains the \pyeobnr~metadata for reference.
In this section we present in detail the architecture and submodules of \pyeobnr.

\pyeobnr~ has been developed with two main objectives. 
The first and foremost is to provide an interface for generating state-of-the-art multipolar waveform models to be used in scientific analyses. 
This functionality can be readily employed for a number of studies and projects related to analysis of GW data.
The second objective is to create a new framework to rapidly develop new EOB waveform models. 
For instance, \pyeobnr~ provides tools such as automatic code generation from \texttt{Mathematica}~\cite{mathematica} files to incorporate the latest theoretical results, a pipeline to calibrate EOB models against NR simulations as well as many others. 
Future \texttt{SEOBNR} models will be developed within this framework; moreover newly developed theoretical results could easily be accompanied by corresponding releases of the \pyeobnr~ package for timely updates to existing models.

\pyeobnr~ has been designed in a modular fashion in order to make sure that it can be easily extended, developed, and maintained in the future. 
The codebase has been separated into a number of submodules, each with a clear designation (see Section~\ref{sec:submodules} for further information on the submodules). 
Detailed and informative documentation accompanies all parts of the code. 
Pre-commit hooks and continuous integration testing are used for code styling, unit and regression tests.

\begin{figure}[t]
\includegraphics[scale=1]{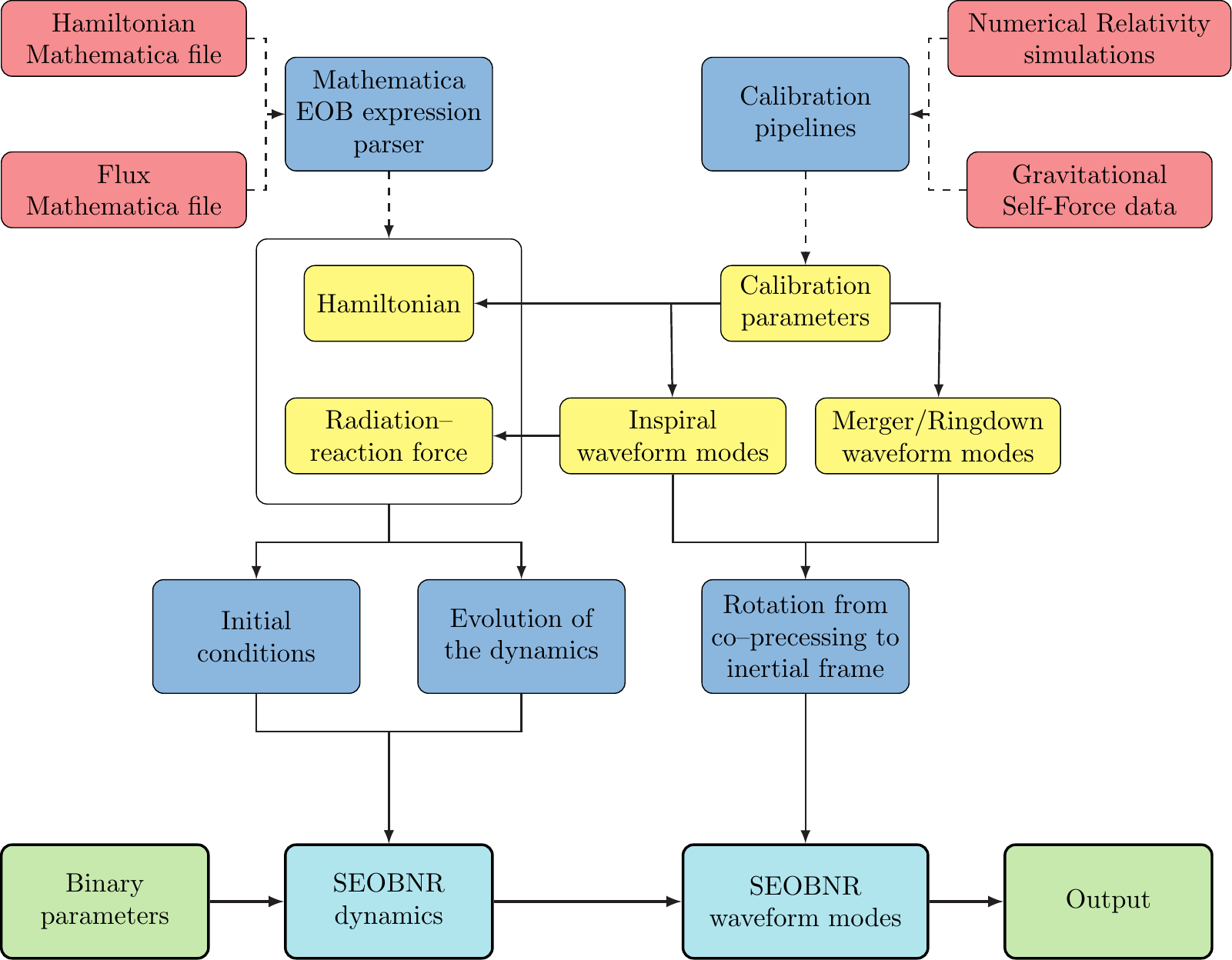}
\caption{Diagram of how the separate \pyeobnr~submodules work and interact with each other. The boxes on the bottom row denote the main sequence of operations executed when generating a (time-domain) waveform. The solid arrows describe how submodules interact to build parts of the waveform, and the dashed lines denote operations which are executed off-line during development and calibration of a new waveform model from scratch.}
\label{fig:code_diagram}
\end{figure}

Fig.~\ref{fig:code_diagram} depicts the overall structure of the \pyeobnr~ package. 
The bottom-most panels show the primary steps involved in generating aligned- and precessing-spin models, while the panels above them provide details on these operations. 
The top-most panels showcase the development functionality, including a code parser for analytical information from \texttt{Mathematica} and a calibration pipeline to Numerical Relativity and Gravitational Self-Force data. 
The remaining parts of the diagram show the important steps in construction of EOB models: the parts of the diagram connected by dashed arrows are carried out off-line, during the development process, while the solid lines show the process of generating a waveform model. 
In physical terms, the dynamics of the binary is computed by numerically integrating the EOB Hamiltonian system with a dissipative radiation-reaction (RR) force that represents the energy loss through radiation of GWs. 
Subsequently, spherical harmonic modes of the emitted waveform are computed in an inertial reference frame, while taking into account how the free parameters of the EOB model have been calibrated against numerical relativity simulations. 
Finally, the two GW polarizations are computed from the waveform modes.

\subsection{Software architecture}

\noindent
The principal functionality of \pyeobnr~ is structured around a single class object \texttt{GenerateWaveform} and a small number of functions used to generate multipolar gravitational waveforms. 
The user is able to specify input parameters through options for initialising the class object and can then obtain output through instance methods. 
In this section we acquaint the reader with the specifics of this class; the variety of options are explained below. 
For examples of their usage, please refer to Section~\ref{sec:UseCases}.

\subsubsection{The GenerateWaveform class}
\label{sec:GWclass}

\noindent
One of the two primary ways to generate a waveform using \pyeobnr~is to make use of the \texttt{GenerateWaveform} class and its instance methods. 
The inputs are provided in a form of a Python dictionary to ensure they can be easily extended. 
There are 27 possible input parameters, only 2 of them are required (the masses \(m_{1}, m_{2}\) of the primary and secondary binary components), while the remaining ones are optional and have default values that are appropriate for most uses, see Table~\ref{table:GWinputs} for details. 
Of the optional parameters:
\begin{itemize}
\item The luminosity \texttt{distance} to the GW source \(d_{L}\) is expressed in units of Mpc, with a default value of 100 (the choice of units is owing to the fact that we dealing with extra-galactic sources).
\item The \texttt{inclination} angle \(\iota\) is the angle between the line-of-sight and the orbital angular momentum in radians. By default, the binary is viewed head-on.
\item The reference frequency \texttt{f\_ref} is significant for the precessing model, where other quantities (e.g. the spins, or the orbital phase) could be defined at a frequency different from the starting frequency. By default it is equal to \texttt{f22\_start}.
\item The orbital phase (at the reference frequency) \texttt{phi\_ref} is measured in radians and is 0 by default.
\item The time spacing \texttt{deltaT} of the waveform is measured in seconds, with a default value of \(1 / 2048\) (used for time-domain outputs and for checking consistency with the Nyquist criterion).
\item The maximum frequency \texttt{f\_max} of the waveform is \SI{1024}{\hertz} by default. If not provided explicitly, its value is computed from \texttt{deltaT} as the Nyquist frequency.
\item The frequency spacing \texttt{deltaF} of the waveform is measured in \SI{}{\hertz}, with a default value of \(0.125\) (used for Fourier-domain outputs).
\item The \texttt{mode\_array} parameter specifies which modes would be computed and returned as output, \emph{in the coprecessing frame}. Only positive\(-m\) modes need to be specified, e.g. \texttt{[(3, 2)]} includes both the \((3, 2)\) and the \((3, -2)\) modes. By default all modes with \(\ell \leq 4\) are selected; at present this includes \((\ell, m) = \{(2, 2), (3, 3), (3, 2), (4, 4), (4, 3)\}\). In addition, for the models described in this publication, the \((\ell, m) = (5, 5)\) mode is available for generation, but is not selected by default.

\item \texttt{approximant} directs which model is used to construct the waveform. At present, \pyeobnr~ includes the \texttt{SEOBNRv5HM} and \texttt{SEOBNRv5PHM} models, while in the future this list will be extended with newly added approximants.

\item The \texttt{conditioning} parameter controls the tapering method which would be applied to the waveform before it is returned. Value \texttt{1} will taper the beginning of the waveform, as it was done for the \texttt{SEOBNRv4PHM} model in \texttt{SimInspiralFD()}. Value \texttt{2} (default) will trigger the standard \texttt{SimInspiralFD()} procedure of adding extra time at the beginning for the purposes of tapering.
\item The \texttt{polarizations\_from\_coprec} option triggers a more efficient computation of the waveform polarizations for the precessing-spin model \texttt{SEOBNRv5PHM} (see Section III C of Ref.~\cite{RamosBuadesv5} for details on this feature).
\item The \texttt{initial\_conditions} parameter controls whether the model would use \texttt{"adiabatic"} (default) or \texttt{"postadiabatic"} initial conditions (described in further detail in Section~\ref{sec:dynamics}). If \texttt{"postadiabatic"} is selected, the option \texttt{initial\_conditions\_postadiabatic\_type} controls whether the \texttt{"analytic"} (default) or \texttt{"numeric"} post-adiabatic regime is applied for computing the initial conditions.
\item Similarly, \texttt{postadiabatic} controls whether the post-adiabatic approximation is used for the solution of the binary inspiral. By default, it is \texttt{True}. See Section~\ref{sec:dynamics} for details.
\end{itemize}
In addition, automatic validation of the input parameters is performed at initialization. These include checks that the required parameters (i.e. the component masses) are correctly provided and that the values of the optional parameters are not outside of their allowed values (e.g. the aligned-spin waveform approximant cannot be executed with non-aligned spins).

Once the \texttt{GenerateWaveform} class has been initialised, the user can execute the waveform generator and obtain output through some of the built-in instance methods. 
Here, it is important to note that the \texttt{GenerateWaveform} class has been designed to be compatible with the standard data analysis package, the LSC's Algorithm Library Suite~(\texttt{LALSuite}) \cite{lalsuite}. 
Therefore, within it we reuse standard \texttt{LALSuite} data types and mimic inputs and outputs of main API functions such as \texttt{SimInspiralChooseTDWaveform()} and its Fourier-domain counterpart, which are provided through the \texttt{lalsimulation} sub-package. 
The available outputs of the waveform generator routine are listed in Table~\ref{table:GWmethods}:
\begin{itemize}
\item \emph{time-domain modes}: obtained using the \texttt{generate\_td\_modes()} method, it produces an array (containing the time domain), and a dictionary containing each of the requested waveform modes (provided through the \texttt{mode\_array} input parameter)
\item \emph{time-domain polarizations}: obtained using the \texttt{generate\_td\_polarizations()} method, it will produce two \texttt{REAL8TimeSeries} containing the plus and cross polarization modes of the waveform
\item \emph{Fourier-domain polarizations}: obtained using the \texttt{generate\_fd\_polarizations()} method, it will produce the Fourier transforms of the plus and cross polarization modes as \texttt{COMPLEX16FrequencySeries}.
\end{itemize}
The output of these class methods are all in SI units: the time domain is provided in seconds, the frequency in Hertz, and the amplitudes of the modes are measured in meters. 
In Section~\ref{sec:UseCases} we provide code examples for using this functionality.

\begin{sidewaystable}
\begin{tabular}{ l l r l }
\toprule
 \textbf{Name} & \textbf{Description} & \textbf{Default value} \\ 
\toprule
 \texttt{mass1} & Mass of companion 1, in solar masses* & N/A \\
\midrule
 \texttt{mass2} & Mass of companion 2, in solar masses* & N/A \\
\midrule
 \texttt{spin1x} & \(x\)-component of dimensionless spin of companion 1 & 0 \\
\midrule
 \texttt{spin1y} & \(y\)-component of dimensionless spin of companion 1 & 0 \\
\midrule
 \texttt{spin1z} & \(z\)-component of dimensionless spin of companion 1 & 0 \\
\midrule
 \texttt{spin2x} & \(x\)-component of dimensionless spin of companion 2 & 0 \\
\midrule
 \texttt{spin2y} & \(y\)-component of dimensionless spin of companion 2 & 0 \\
\midrule
 \texttt{spin2z} & \(z\)-component of dimensionless spin of companion 2 & 0 \\
\midrule
 \texttt{distance} & Distance to the source, in Mpc & 100 \\
\midrule
 \texttt{inclination} & Inclination of the source, in radians & 0 \\
\midrule
 \texttt{f22\_start} & Starting waveform generation frequency, in Hz & 20 \\
\midrule
 \texttt{f\_ref} & The reference frequency, in Hz & \texttt{f22\_start} \\
\midrule
 \texttt{phi\_ref} & Orbital phase at the reference frequency, in radians & 0 \\
\midrule
 \texttt{deltaT} & Time spacing, in seconds & \(1/2048\) \\
\midrule
 \texttt{f\_max} & Maximum frequency, in Hz & 1024 Hz \\
\midrule
 \texttt{deltaF} & Frequency spacing, in Hz & 0.125 \\
\midrule
 \texttt{mode\_array} & Mode content\(^{\dagger}\) & \texttt{None} (i.e. all modes with \(\ell \leq 4\)) \\ 
\midrule
 \texttt{approximant} & Name of the waveform approximant to be used & \texttt{{\qq}SEOBNRv5HM{\qq}} \\
\midrule
 \texttt{conditioning} & Conditioning procedure for the waveform & 2 \\
\midrule
 \texttt{polarizations\_from\_coprec} & Whether to generate the polarizations from the co-precessing frame modes & \texttt{True} \\
\midrule
 \texttt{initial\_conditions} & Mode of initial conditions to be used (adiabatic or post-adiabatic) & \texttt{"adiabatic"} \\
\midrule
 \texttt{initial\_conditions\_postadiabatic\_type} & Type of post-adiabatic initial conditions & \texttt{"analytic"} \\
\midrule
 \texttt{postadiabatic} & Whether to use the post-adiabatic approximation for the inspiral & \texttt{True} \\
\midrule
 \texttt{postadiabatic\_type} & Type of post-adiabatic inspiral & \texttt{"analytic"} \\
\bottomrule
\end{tabular}
\caption{Input parameters of the \texttt{GenerateWaveform} class. The masses of the primary and secondary components (marked with *) are always required. When specifying the mode content of the output (marked with \(^{\dagger}\)), only the positive modes need to be specified, e.g \texttt{[(2,2),(2,1)]}. In the future, additional waveform approximant names will be added to this interface as they become part of the \pyeobnr~package.}
\label{table:GWinputs}
\end{sidewaystable}

\subsubsection{Advanced interface: Generating modes and polarizations directly}

\noindent
Expert users who seek  to perform waveform validation have access to a set of specialized functions which can be used to generate the waveform modes or the waveform polarizations.

The function \texttt{generate\_modes\_opt()} may be used for both aligned- and precessing-spin waveforms. For this function, both inputs and outputs are in geometric units. In Section~\ref{sec:UseCases} we provide an example of how to use this endpoint, together with an explanation of its inputs and outputs.

In addition, for precessing waveform models, we provide the function \texttt{generate\_prec\_hpc\_opt()} that directly generates the two GW polarizations from the waveform modes in the co-precessing frame, which tracks the motion of the orbital plane in a precessing binary. This function bypasses the construction of the inertial frame modes. This makes the waveform generation significantly faster, especially in the case of low total mass binaries. The user must provide values for the inclination and phase parameters. Since this is significantly faster than the canonical way of generating precessing-spin waveforms, it is also the default approach used for parameter estimation analyses.

\subsection{Primary submodules}
\label{sec:submodules}
\noindent
In this section we describe the primary submodules responsible for generating waveforms. In particular, we consider modules which contain theoretical EOB expressions that form building blocks of the \texttt{SEOBNRv5} models, as well as the infrastructure which allows us to compute the binary inspiral from the Hamiltonian and the RR force, compute the waveform modes, and finally attach the merger-ringdown modes to form the complete IMR waveform.

\subsubsection{Hamiltonian}

\noindent
Consider a binary with masses \(m_{1}\) and \(m_{2}\) (\(m_{1} \geq m_{2}\)) and spins \(\mathbf{S}_{1}\) and \(\mathbf{S}_{2}\). The EOB formalism relies on an effective Hamiltonian (\(H_{\mathrm{eff}}\)) of a test mass \(\mu = m_{1} m_{2} / (m_{1} + m_{2})\) moving in the Kerr space-time of a body with mass \(M = m_{1} + m_{2}\), with deformation parameter \(\nu = \mu / M\). The conservative two-body dynamics is obtained from the EOB Hamiltonian:
\begin{align}
H_{\mathrm{EOB}} = M \sqrt{1 + 2 \nu \left(\frac{H_{\mathrm{eff}}}{\mu} - 1\right)}.
\end{align}
The dynamical variables on which the value of the Hamiltonian depends are the orbital separation \(\mathbf{r}\), the conjugate momentum \(\mathbf{p}\), and the spins \(\mathbf{S}_{1,2}\). 
In addition, its value depends on a set of calibration parameters \(\mathbf{\theta}\), that correspond to yet unknown higher-order PN coefficients. In this section we describe how the EOB Hamiltonian is implemented on \pyeobnr.

The \texttt{hamiltonian} submodule contains the entire infrastructure for using EOB Hamiltonians. 
Analytical expressions are converted from \texttt{Mathematica} files into \texttt{cython} source files which subclass the \texttt{Hamiltonian\_C} abstract class (or \texttt{Hamiltonian\_v5PHM\_C} for the \texttt{SEOBNRv5PHM} precessing-spin model). These files provide the infrastructure to evaluate the EOB Hamiltonian, as well as its Jacobian and Hessian (the derivatives are computed automatically during the conversion from \texttt{Mathematica} to \texttt{cython} using \texttt{wolframclient} as these are more computationally efficient compared to either finite-difference derivatives or automatic derivatives using the \texttt{jax} \cite{jax2018github} package). In addition, further quantities derived from the EOB Hamiltonian are provided through methods of this class: \texttt{xi} computes the tortoise coordinate conversion factor \(\xi(r) = p_{r_*} / p_r = \mathrm{d}r / \mathrm{d}r_*\); \texttt{auxderivs} provides access to some of the potentials computed as part of computing the value of \(H_{\mathrm{EOB}}\), which are necessary for optimising the computational efficiency of the post-adiabatic approximation by using analytic derivatives (see Section~\ref{sec:dynamics} for details).

The infrastructure provided in \pyeobnr will allow for an easy extension of the current Hamiltonians to include more analytical information. 
In future releases of \pyeobnr, the \texttt{hamiltonian} submodule will be extended in order to include additional physical effects in the Hamiltonian, for instance tidal corrections, needed for generating waveforms describing binary neutron-star coalescences.

\begin{figure}[t]
    \includegraphics[scale=1]{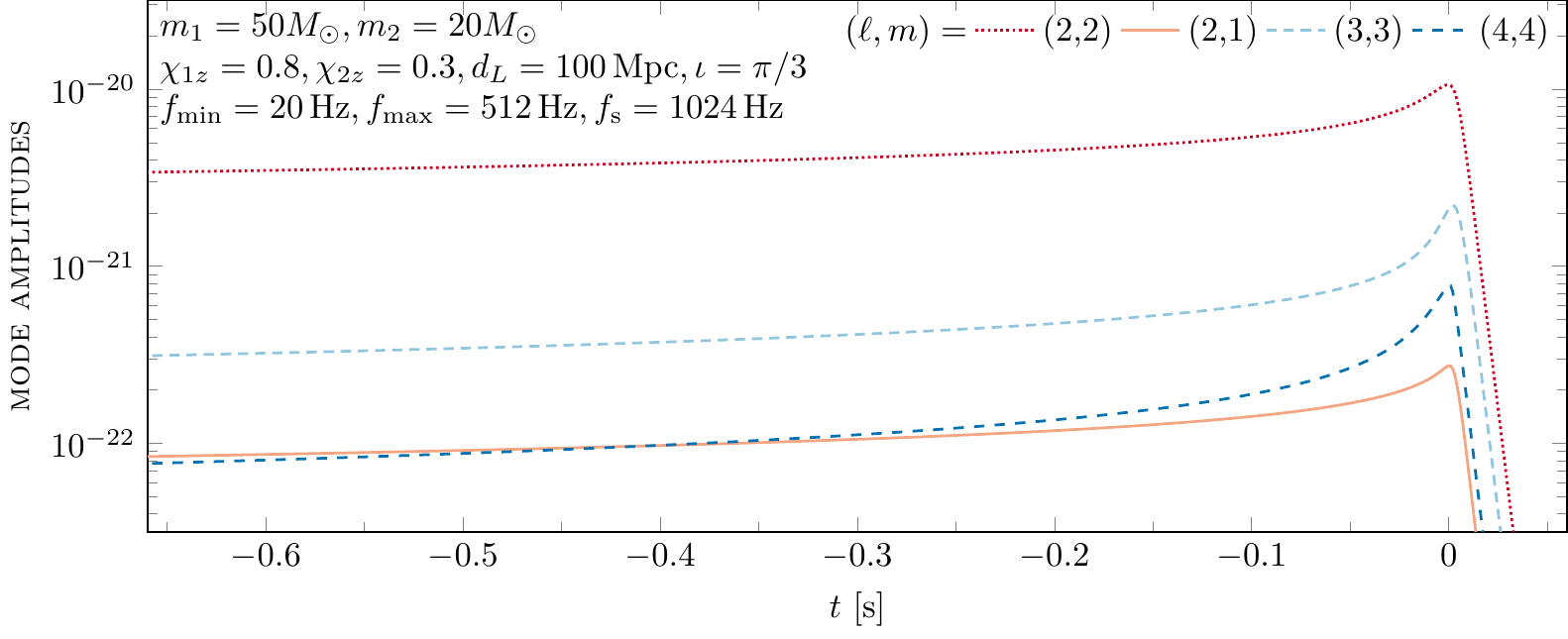}
    \caption{Plot of the waveform modes produced with the \texttt{generate\_td\_modes()} method for an aligned-spin black-hole binary. The parameters used for waveform generation are stated in the figure, with the sampling rate \(f_{\mathrm{s}} = 1 / \text{\texttt{deltaT}}\).}
    \label{fig:td_modes}
\end{figure}

\subsubsection{Waveform and radiation-reaction force}

\noindent
Like the Hamiltonian, RR force is also a fundamental building block of the \texttt{SEOBNR} dynamics. In \pyeobnr, the RR force is computed as an iterative sum over factorized/resummed PN expressions for the waveform modes \citep{Damour:2008gu, Pan:2010hz}:
\begin{align}
\mathbf{\mathcal{F}} = \frac{M \Omega}{16 \pi} \frac{\mathbf{p}}{L} \sum_{\ell = 2}^{L_{max}} \sum_{m = - \ell}^{\ell} m^{2} |d_{L} h_{\ell m}|^{2},
\end{align}
where \(\Omega\) is the angular orbital frequency, \(L\) is the magnitude of the orbital angular momentum, \(\mathbf{p}\) is the canonical momentum, \(d_{L}\) is the luminosity distance and \(h_{\ell m}\) are the gravitational modes far from the source. Due to the large number of terms (in \texttt{SEOBNRv5HM} and \texttt{SEOBNRv5PHM}, \(L_{max} = 8\) is used), the current implementation of the RR force is one area where significant improvements in efficiency could be achieved, which in turn will have an effect on the overall speed of the model.

\subsubsection{Dynamics}
\label{sec:dynamics}

\noindent
For a binary system with aligned or anti-aligned spins, the dynamics is obtained by solving the Hamilton equations:
\begin{subequations}
\begin{align}
\dot{\mathbf{r}} &= \frac{\partial H_{\mathrm{EOB}}}{\partial \mathbf{p}} \\
\dot{\mathbf{p}} &= - \frac{\partial H_{\mathrm{EOB}}}{\partial \mathbf{r}} + \mathbf{\mathcal{F}}
\end{align}
\end{subequations}
For a generic-spin system, there are further 6 equations of motion governing the spin evolution:
\begin{align}
\dot{\mathbf{S}}_{1, 2} = \frac{\partial H_{\mathrm{EOB}}}{\partial \mathbf{S}_{1, 2}} \times \mathbf{S}_{1, 2}.
\end{align}

Interestingly, the dynamics may be approximated by a post-adiabatic approach which allows us to compute the evolution of the orbital parameters on a sparse grid in radial domain~\cite{Nagar:2018gnk, Rettegno:2019tzh, Gamba:2021ydi, Mihaylov:2021bpf}. 
The approximation is achieved iteratively and is valid until a short time before merger. The approach requires us to find the solutions of (non-linear) algebraic equations in order to approximate the conjugate momentum to the tortoise radial coordinate \(p_{r_{*}}\) and the orbital angular momentum \(p_{\varphi}\):
\begin{subequations}
\begin{align}
\frac{\dd p_{\varphi}}{\dd r} \frac{\partial H_{\mathrm{EOB}}}{\partial p_{r_{*}}} - \mathcal{F}_{\varphi} &= 0 \\
\frac{\partial H_{\mathrm{EOB}}}{\partial p_{r_{*}}} + \frac{\partial H_{\mathrm{EOB}}}{\partial r} \frac{\dd r}{\dd p_{r_{*}}} - \frac{p_{r_{*}}}{p_{\varphi}}\,\mathcal{F}_{\varphi} &= 0 
\end{align}
\end{subequations}
The \texttt{dynamics} module provides the functionality for computing the binary dynamics of the EOB models. In order to compute the inspiral dynamics up to merger, the code starts by computing the initial conditions for the binary from the user input parameters. The dynamics computation  can be performed either by integrating the Hamilton equations numerically, using standard ODE integrators,  or alternatively by computing the post-adiabatic approximation to the dynamics, which is more efficient and is therefore the default method in the precessing model \texttt{SEOBNRv5PHM}. It should be noted that the post-adiabatic approximation cannot be extended to the entire waveform and therefore the final few orbits need to be computed using the ODE integrator, with the results from both computations merged to provide the full dynamics.

\subsubsection{Fits}

\noindent
In the previous section we have outlined that in order to achieve the desired accuracy of the models, data from NR simulations as well as data from gravitational self-force computations is used to calibrate the model. While the details of the calibration process are presented in Section~\ref{sec:calibration}, here we discuss how the output of this calibration process is stored and utilised in waveform generation. The calibration pipelines find values of the free parameters of the model~\cite{Pompiliv5} which provide the best match between the \texttt{SEOBNR} models and the NR simulations. The values of these free parameters are computed, however, at discrete values of the binary parameters (since NR simulations are not available everywhere in parameter space). In order to provide coverage for the entire parameter range of the \texttt{SEOBNR} models, the values of these free parameters are fitted using a least-squares method, and the coefficients of these interpolation fits are stored as part of the Fits submodule. In addition, this submodule provides an interface which is used by the dynamics module in order to compute the values of the coefficients when the dynamics is generated. Currently, the fits are implemented as  polynomial or radical functions of the physical variables (for details please refer to Appendices A, B, C, and D of Ref.~\cite{Pompiliv5}).

\begin{center}
    \begin{table}[b]
    \begin{tabular}{ p{6.5cm} p{7cm} c }
    \toprule
     \textbf{Method} & \textbf{LALSuite counterpart} & \textbf{Output} \\ 
    \toprule
     \multirow{ 2}{*}{\texttt{generate\_td\_modes()}} & \multirow{ 2}{*}{\texttt{SimInspiralChooseTDModes()}} & \texttt{times}, \\
     & & \texttt{hlm\_dict} \\
    \midrule
     \multirow{ 2}{*}{\texttt{generate\_td\_polarizations()}} & \multirow{ 2}{*}{\texttt{SimInspiralChooseTDWaveform()}} & \texttt{hp}, \\
     & & \texttt{hc} \\
    \midrule
     \multirow{ 2}{*}{\texttt{generate\_fd\_polarizations()}} & \multirow{ 2}{*}{\texttt{SimInspiralChooseFDWaveform()}} & \texttt{hptilde}, \\
     & & \texttt{hctilde} \\
    \bottomrule
    \end{tabular}
    \caption{Output of the three separate instance methods of the \texttt{GenerateWaveform} class for producing the gravitational waveform modes or polarizations. Data types \texttt{REAL8TimeSeries} and \texttt{COMPLEX16FrequencySeries} follow \texttt{LALSuite} conventions~\cite{lalsuite}.}
    \label{table:GWmethods}
    \end{table}
\end{center}

In the future, more sophisticated methods will be implemented for utilising the NR and GSF data. In particular, infrastructure and methods for dealing with the uncertainties in the computed values of the free parameters will provide better agreement with NR simulations, and could allow us to translate uncertainties in the provided data into uncertainties of the waveform model itself.

\begin{figure}[t]
    \includegraphics[scale=1]{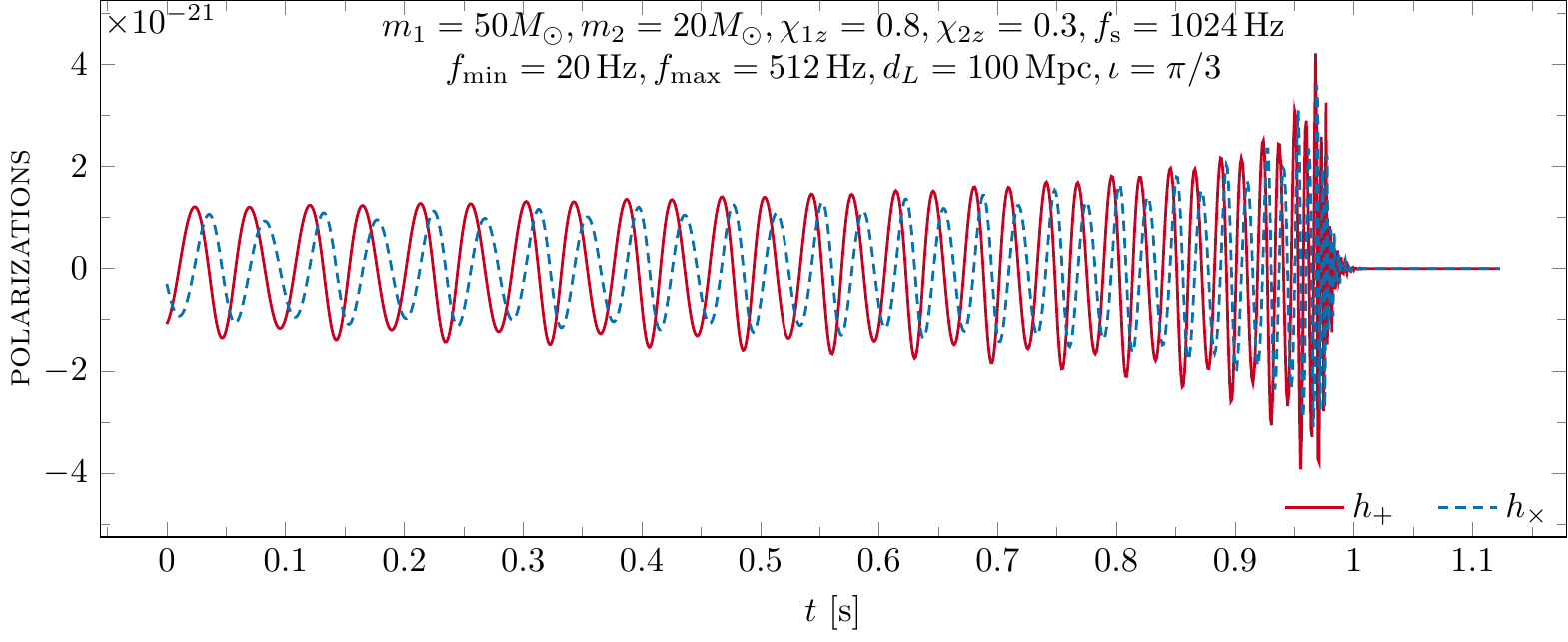}
    \caption{Plot of the waveform polarizations in time domain produced with the \texttt{generate\_td\_polarizations()} method. The parameters used for waveform generation are stated in the figure, with the sampling rate \(f_{\mathrm{s}} = 1 / \text{\texttt{deltaT}}\).}
    \label{fig:td_polarizations}
\end{figure}

\subsection{Auxiliary submodules}

\noindent
In addition to the main submodules needed to generate the waveform, a development-oriented extension to \pyeobnr~ also provides a set of tools whose purpose is to help with the \emph{development} of new waveform models. While these are aimed at advanced users of the package who would like to either build new models or provide improvements to current ones, we give an outline of them here. The most important development tools are the Mathematica expression parser which translates analytical EOB expressions into usable \texttt{python/cython} code, and the calibration pipelines which allow us to constrain free parameters in the model to achieve better agreement with NR waveforms. Please note that unlike the much more commonly used waveform generation tools, these modules have not passed a software review process.

\subsubsection{Automatic code generation}

\noindent
\pyeobnr~contains a parser which can be used to translate \texttt{Mathematica} code into python and performant \texttt{cython} code leveraging the capabilities of the \texttt{sympy} package~\cite{10.7717/peerj-cs.103},  which can then be used as part of the Hamiltonian and RR force modules. Mathematica files containing analytical EOB Hamiltonian expressions can be parsed using the following command

\texttt{python generate\_Hamiltonian.py --ham\_file [path to Hamiltonian file] --name [name for this Hamiltonian]}

The variables of the analytical expressions will be automatically parsed; the variables, constants, and calibration parameters will be identified and translated in the correct fashion to \texttt{cython}. The output file is ready to be directly used by the dynamics module without any further preparation. The code generation can be fine-tuned by supplying additional options via \texttt{toml} \cite{toml, pytoml} files. In the future, the options provided in this parser will need to be extended in order to cover additional parameters in the analytical Hamiltonians which will be used in future \texttt{SEOBNR} waveform models.

\subsubsection{Calibration pipelines}
\label{sec:calibration}

\noindent
In order to constrain the free coefficients in \texttt{SEOBNR} waveform models, we first need to establish a metric for comparison between two waveforms. This is provided through the \texttt{metrics} class. Several different metrics are provided,   for example:  the mismatch between NR and \texttt{SEOBNR} waveforms~\cite{PhysRevD.94.024012},  the difference in time to merger between the two waveforms.

The parameter search and estimation is done using Bayesian stochastic sampling,  performed via the functionality in the \texttt{bilby} package,   which is commonly used for data analysis in the GW community.  In particular, we use the machine-learning enhanced \texttt{nessai} sampler~\cite{Williams:2021qyt}, which provides very quick convergence for our class of problems. In order to submit a calibration run, a user may use the commands below:

\texttt{python batch\_calibration\_setup.py --cases-file [path to NR catalogue] --prior 4D\_tight\_a6.prior --sampler nessai --scheduler slurm --queue hypatia --singularity --singularity-image [path to .sing file] --calibration-settings calibration\_settings.toml
bash ./submit\_all.sh}

The output of this process, once completed, will be a set of posterior \texttt{JSON} files which can then be used to assign values for each free parameter at each binary parameter point. The interpolations of these parameters are then used by the Fits package in order to provide values for the free parameters of the model during execution.

\begin{figure}[t]
    \includegraphics[scale=1]{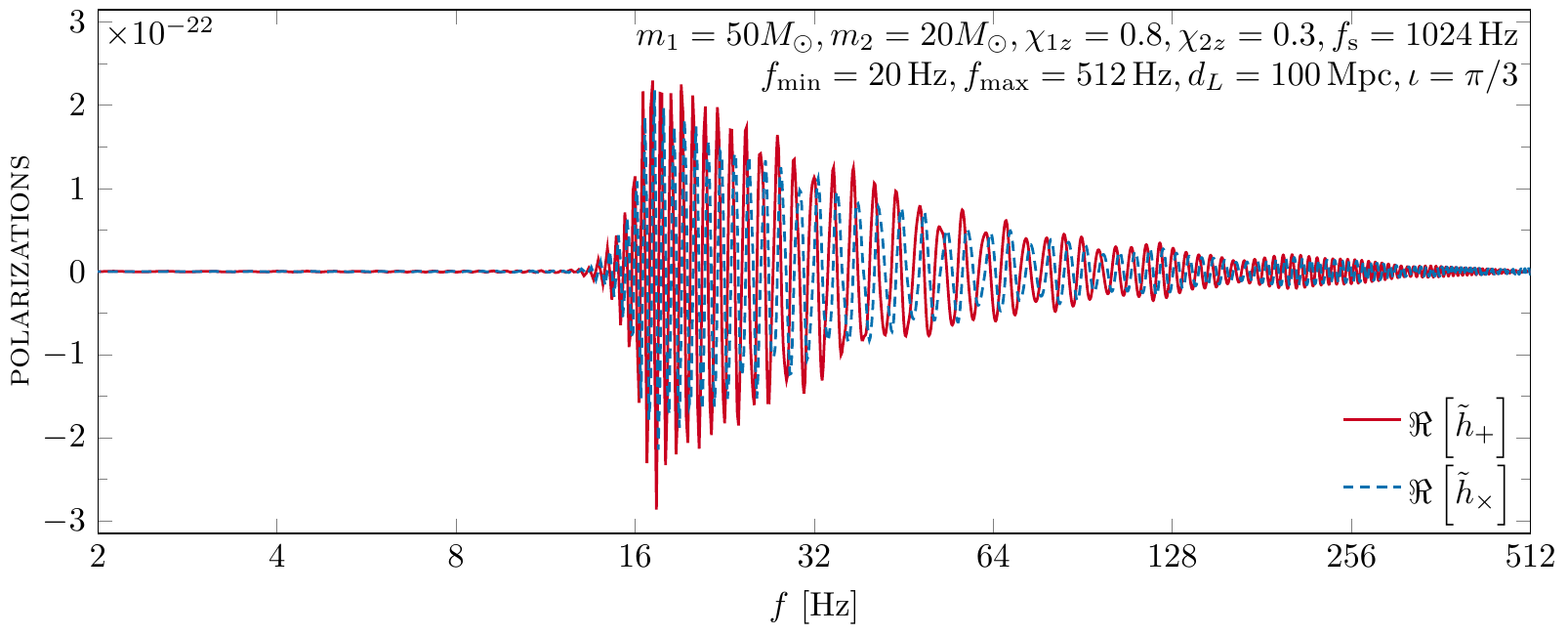}
    \caption{Plot of the waveform polarizations in Fourier domain produced using the \texttt{generate\_fd\_polarizations()} method. The parameters used for waveform generation are stated in the figure, with the sampling rate \(f_{\mathrm{s}} = 1 / \text{\texttt{deltaT}}\).}
    \label{fig:fd_polarizations}
\end{figure}


\section{Use cases}
\label{sec:UseCases}

\noindent
In the previous section, we described in detail the structure of the \pyeobnr~ package. Here, we show how to use the \texttt{SEOBNRv5HM} and \texttt{SEOBNRv5PHM} models to generate waveforms in several common cases.

\subsection{Generating an aligned-spin gravitational waveform}

\noindent
For the case of a binary with aligned (or anti-aligned) spins, we will demonstrate the use of two different mechanisms.  First we consider the most likely use case,  such as the one that occurs during GW parameter estimation - the input and output are in physical (SI) units.
As shown in listing~\ref{lst_std} we use the \texttt{GenerateWaveform} class, which we presented in detail in Section~\ref{sec:GWclass}.  We select a binary of total mass \(M = 80 M_{\odot}\), with a starting frequency of 20 Hz and an appropriate sampling rate of 1024 Hz. The dimensionless spins of the primary and the secondary binary components are 0.8 and 0.3, respectively. The binary is at a distance of 100 megaparsecs and is inclined at an angle of \(\pi / 3\).

After the instance has been initialised, we invoke the methods \texttt{generate\_td\_modes()} and \texttt{generate\_td\_polarizations()} (see Section~\ref{sec:GWclass} and Table~\ref{table:GWmethods}) in order to obtain the time-domain modes and polarizations, respectively. For the binary configuration in question, it takes \SI{106}{\milli\second} to generate the waveform modes, and \SI{113}{\milli\second} to generate the polarizations. Readers who would like to learn more about the efficiency of the aligned-spin model can refer to~\cite{Pompiliv5}.

\begin{minipage}{\linewidth}
\begin{lstlisting}[
	language=Python,
	caption={Generating aligned-spin waveform modes and polarizations in physical units with LAL conventions.},
	label=lst_std,
	xleftmargin=\parindent,
	xrightmargin=0.7cm,
]
import numpy as np
from pyseobnr.generate_waveform import GenerateWaveform

m1, m2 = 50., 20.
s1x, s1y, s1z = 0., 0., 0.8
s2x, s2y, s2z = 0., 0., 0.3

deltaT = 1./1024.
f_min = 20.
f_max = 512.

distance = 100.
inclination = np.pi / 3.
phi_ref = 0.
approximant = "SEOBNRv5HM"

params_dict = {
    "mass1": m1, "mass2": m2,
    "spin1x": s1x, "spin1y": s1y, "spin1z": s1z,
    "spin2x": s2x, "spin2y": s2y, "spin2z": s2z,
    "deltaT": deltaT,
    "f22_start": f_min,
    "phi_ref": phi_ref,
    "distance": distance,
    "inclination": inclination,
    "f_max": f_max,
    "approximant": approximant,
}

waveform_gen = GenerateWaveform(params_dict)
t, hlm = waveform_gen.generate_td_modes()
hp, hc = waveform_gen.generate_td_polarizations()
\end{lstlisting}
\end{minipage}

For waveform development and validation it is frequently helpful to instead work in geometric units, e.g. to ease the comparison with NR results. For this we use the function \texttt{generate\_modes\_opt()}, as shown in listing~\ref{lst_opt}.

\begin{minipage}{\linewidth}
\begin{lstlisting}[
	language=Python,
	caption={Using the internal SEOBNRv5HM aligned-spin model generator to obtain the waveform modes and\\time domain in geometric units.},
	label=lst_opt,
	xleftmargin=\parindent,
	xrightmargin=0.7cm,
]
from pyseobnr.generate_waveform import generate_modes_opt

q = 5.3 # mass ratio
chi_1 = 0.9 # spin of the primary
chi_2 = 0.3 # spin of the secondary
omega0 = 0.0137 # orbital frequency in geometric units with M = 1

t, modes = generate_modes_opt(q, chi_1, chi_2, omega0)
\end{lstlisting}
\end{minipage}

\begin{minipage}{\linewidth}
\begin{lstlisting}[
	language=Python,
	caption={Generating precessing-spin waveform modes and polarizations in physical units with LAL conventions.},
	label=lst_prec_std,
	xleftmargin=\parindent,
	xrightmargin=0.7cm,
]
import numpy as np
from pyseobnr.generate_waveform import GenerateWaveform

m1, m2 = 50., 20.
s1x, s1y, s1z = 0.5, 0., 0.5
s2x, s2y, s2z = 0., 0.5, 0.5

deltaT = 1./1024.
f_min = 20.
f_max = 512.

distance = 1000.
inclination = np.pi / 3.
phi_ref = 0.
approximant = "SEOBNRv5PHM"

params_dict = {
    "mass1": m1, "mass2": m2,
    "spin1x": s1x, "spin1y": s1y, "spin1z": s1z,
    "spin2x": s2x, "spin2y": s2y, "spin2z": s2z,
    "deltaT": deltaT,
    "f22_start": f_min,
    "phi_ref": phi_ref,
    "distance": distance,
    "inclination": inclination,
    "f_max": f_max,
    "approximant": approximant,
}

waveform_gen = GenerateWaveform(params_dict)
t, hlm = waveform_gen.generate_td_modes()
hp, hc = waveform_gen.generate_td_polarizations()
\end{lstlisting}
\end{minipage}

\subsection{Generating a precessing-spin gravitational waveform}

\noindent
We can also use the \texttt{GenerateWaveform} class to produce waveforms from a binary with precessing spins with minimal changes. 
Apart from specifying non-zero \(x\)- and \(y\)-components of the spins, the user needs to specify the precessing-spin \texttt{SEOBNRv5PHM} approximant. 
Readers can use the code in Listing~\ref{lst_prec_std} to generate a precessing-spin waveform. 
For this precessing binary, \pyeobnr~ takes \SI{395}{\milli\second} to generate the waveform modes, and only \SI{387}{\milli\second} to generate the polarizations. Readers who would like to learn more about the efficiency of the precessing-spin model are encouraged to refer to~\cite{RamosBuadesv5}.

The polarizations in this case can be obtained by using the function \texttt{generate\_prec\_hpc\_opt()}, an example of which is shown in listing~\ref{lst_prec}.

\begin{minipage}{\linewidth}
\begin{lstlisting}[
	language=Python,
	caption={Using the \texttt{generate\_prec\_hpc\_opt} endpoint to generate the polarizations from the co-precessing frame modes.},
	label=lst_prec,
	xleftmargin=\parindent,
	xrightmargin=0.7cm,
]
from pyseobnr.generate_waveform import generate_prec_hpc_opt

q = 2.0 # mass ratio
chi_1 = np.array([0.5, 0.0, 0.5]) # spin of the primary
chi_2 = np.array([0.0, 0.5, 0.5]) # spin of the secondary
omega0 = 0.01 # orbital frequency in geometric units with M=1

 _, _, model = generate_prec_hpc_opt(
    q,
    chi_1, chi_2,
    omega0,
    debug=True,
    settings={
        "phi_ref": np.pi / 2,
        "inclination": np.pi / 3,
    },
)
\end{lstlisting}
\end{minipage}

\section{Conclusions and future developments}
\label{sec:future}

\noindent
The \pyeobnr~package provides a flexible and modern infrastructure for developing waveform models in the \texttt{SEOBNR} framework.
This Python-based code has allowed us to significantly cut down research and development time and has reduced the amount of
boiler-plate code required to build, tune, and implement a working EOB model as compared to the legacy development
environment which heavily relied on C/C++ \cite{kernighan2006c, ISO:1998:IIP} and where models where implemented in C99 \cite{ISO:C99} in LALSuite~\cite{lalsuite}.
For the sake of compatibility we have retained some of the necessary data types and main interfaces provided by LALSuite.
In this publication we have discussed the main capabilities and features of the package, and showcased waveform models for binary 
black holes with aligned or precessing spins on quasi-circular orbits.

We are planning to significantly enhance the capabilities of \pyeobnr~in the future. In addition to the quasi-circular  aligned- and precessing-spin model, waveforms for eccentric binaries will be provided in order to aid the community-wide effort to survey eccentricity and unveil the origin of the observed compact-binary populations. 
Moreover, waveforms will be extended for use in tests of General Relativity, which would enable these analyses to be performed using \pyeobnr. 
Additionally, including tidal corrections will allow us to perform inference runs for binary neutron stars and neutron-star--black-hole binaries, which would need to be particularly efficient in the low-mass or large mass-ratio regimes.

Finally, we would like to emphasize that building \pyeobnr~is a step towards ensuring that the development of new waveform models for LVK and future detectors can proceed using the most modern and sophisticated computing tools and methods. We anticipate that the addition of GPU support and the use of neural networks will make future \texttt{SEOBNR} waveform models even more efficient, in turn enabling us to pursue outstanding new science in gravity, fundamental physics, cosmology and astrophysics.

\section*{Acknowledgements}

The authors thank sincerely the LVK team responsible for the review of \pyeobnr~ and the \texttt{SEOBNRv5} models: Geraint Pratten, Stanislav Babak, Alice Bonino, Eleanor Hamilton, N. V. Krishnendu, Piero Rettegno, Riccardo Sturani, and Jooheon Yoo. The development work for this software package was carried out on the \texttt{Hypatia} computing cluster at the Max Planck Institute for Gravitational Physics in Potsdam, Germany.

This research has made use of data or software obtained from the Gravitational Wave Open Science Center (gwosc.org), a service of LIGO Laboratory, the LIGO Scientific Collaboration, the Virgo Collaboration, and KAGRA. 
LIGO Laboratory and Advanced LIGO are funded by the United States National Science Foundation (NSF) as well as the Science and Technology Facilities Council (STFC) of the United Kingdom, the Max-Planck-Society (MPS), 
and the State of Niedersachsen/Germany for support of the construction of Advanced LIGO and construction and operation of the GEO600 detector. Additional support for Advanced LIGO was provided by the Australian Research Council. 
Virgo is funded, through the European Gravitational Observatory (EGO), by the French Centre National de Recherche Scientifique (CNRS), the Italian Istituto Nazionale di Fisica Nucleare (INFN) and the Dutch Nikhef, with contributions 
by institutions from Belgium, Germany, Greece, Hungary, Ireland, Japan, Monaco, Poland, Portugal, Spain. KAGRA is supported by Ministry of Education, Culture, Sports, Science and Technology (MEXT), Japan Society for the Promotion of 
Science (JSPS) in Japan; National Research Foundation (NRF) and Ministry of Science and ICT (MSIT) in Korea; Academia Sinica (AS) and National Science and Technology Council (NSTC) in Taiwan.

\begin{appendices}
\section{Code for generating the plots of the waveform}

\noindent
In order to aid users in getting started and actively using \pyeobnr, in Listing~\ref{lst_figs} we provide the code for generating the figures of the waveform mode amplitudes (Fig.~\ref{fig:td_modes}), the time-domain polarizations (Fig.~\ref{fig:td_polarizations}), and the Fourier-domain polarizations (Fig.~\ref{fig:fd_polarizations}).

\begin{minipage}{\linewidth}
\begin{lstlisting}[
	language=Python,
	caption={Generating the plots appearing in Figs.~\ref{fig:td_modes}, \ref{fig:td_polarizations}, and \ref{fig:fd_polarizations}.},
	label=lst_figs,
	xleftmargin=\parindent,
	xrightmargin=0.7cm,
]
import matplotlib.pyplot as plt

plt.figure()

# Create the plot in Figure 2
for mode in [(2, 2), (2, 1), (3, 3), (4, 4)]:
    plt.plot(t, np.abs(hlm[mode]))
    
plt.yscale('log')
plt.ylim(1e-24, 1e-19)
plt.savefig("figure2.png")

# Create the plot in Figure 3
plt.figure()

plt.plot(t, hp.data.data)
plt.plot(t, hc.data.data)

plt.savefig("figure3.png")

# Create the plot in Figure 4
plt.figure()

plt.plot(f, np.real(hp_f.data.data))
plt.plot(f, np.real(hc_f.data.data))

plt.savefig("figure4.png")
\end{lstlisting}
\end{minipage}
\end{appendices}

\bibliographystyle{apsrev4-2}
\bibliography{SEOBNRv5_software}

\end{document}